\documentclass[preprintnumbers,amsmath,11pt,amssymb,floatfix,
superscriptaddress,nofootinbib,showpacs]{revtex4}
\usepackage{graphicx}
\usepackage{epsfig}
\usepackage{bm}
\usepackage{amsfonts}

\input epsf

\begin{document}

\title{\Large A study of generalized second law of thermodynamics in modified $f(R)$ Horava-Lifshitz gravity}

\author{\bf Surajit Chattopadhyay}
\email{surajit_2008@yahoo.co.in, surajcha@iucaa.ernet.in}
\affiliation{Pailan College of Management and Technology, Bengal
Pailan Park, Kolkata-700 104, India.}

\author{\bf Rahul Ghosh}
\email{ghoshrahul3@gmail.com} \affiliation{Department of
Mathematics, Bhairab Ganguly College, Kolkata-700 056, India.}

\date{\today}

\begin{abstract}
This work investigates the validity of the generalized second law
of thermodynamics in modified $f(R)$ Horava-Lifshitz gravity
proposed by Chaichian et al (2010) [\emph{Class. Quantum Grav.} 27
(2010) 185021], which is invariant under foliation-preserving
diffeomorphisms. It has been observed that the equation of state
parameter behaves like quintessence $(w>-1)$. We study the
thermodynamics of the apparent, event and particle horizons in
this modified gravity. We observe that under this gravity, the
time derivative of total entropy stays at positive level and hence
the generalized second law is validated.
\end{abstract}


\maketitle

\section{\bf{Introduction}}
It is suggested in the modified gravity approach that the
accelerated expansion of the universe \cite{Riess},
\cite{padmanabhan} is caused by a modification of gravity at the
early/late-time universe. For reviews of the modified gravity the
readers are suggested to see \cite{Nojiri}. Application of
Horava–Lifshitz gravity as a cosmological framework has given rise
to Horava–Lifshitz cosmology, which has proven to lead to
interesting behaviors \cite{Saridakis}. In a recent work,
reference \cite{Chaichian} proposed a general modified
Horava-Lifshitz gravity that could be easily related to $f(R)$,
which is a traditional modified theory of gravity. This was termed
as modified $f(R)$ Horava-Lifshitz gravity, that would be
abbreviated as MFRHL in the remaining part of the paper. This has
been shown in \cite{Chaichian} that for a special choice of
parameters the MFRHL coincides with the traditional $f(R)$-gravity
on the spatially flat Friedman-Robertson-Walker (FRW) background.
For a standard $f(R)$ gravity, the action is given by
\cite{Chaichian}
\begin{equation}
S_{f(R)}=\int d^{4}x\sqrt{-g}f(R)
\end{equation}
where, $f(R)$ is a function of the scalar curvature $R$. Detailed
discussion on the mathematical background of $f(R)$ gravity is
given in the references \cite{Gourgoulhon}. In the standard $f(R)$
gravity, the metric is given by \cite{Chaichian}
\begin{equation}
d{s^2}=-N^2dt^2+g^{(3)}_{ij}(dx^i +N^idt)(dx^j+N^jdt),
~~~~~~~~~~~~i=1,2,3
\end{equation}
Here $N$is called the lapse variable and $N^i$'s are the shift
variables. Then the scalar curvature $R$ has the form
\cite{Chaichian}
\begin{equation}
R=K^{ij}K_{ij}-K^2+R^{(3)}+2\nabla_{\mu} (n^\mu\nabla_\nu
n^\nu-n^\nu\nabla_{\nu} n^\mu)
\end{equation}
and $\sqrt{-g}= \sqrt{g^{(3)}}N$. Here $R^{(3)}$ is the
three-dimensional scalar curvature defined by the metric $g^{(3)}$
and $K_{ij}$ is the extrinsic curvature defined by
\begin{equation}
K_{ij}=\frac{1}{2N}(\dot{g}^{(3)}_{ij}-\nabla^{(3)}_{i}
N_j-\nabla^{(3)}_{j} N_{i})~~~~~~~~~~K=K^{i}_{j}
\end{equation}
$n^\mu$ is a unit vector perpendicular to the three-dimensional
hypersurface $\Sigma_t$ defined by $t$= constant and
$\nabla^{(3)}_{i}$ expresses the covariant derivative on the
hypersurface $\Sigma_{t}$. Details of the mathematical background
of $f(R)$ gravity is available in the references \cite{Kluson} and
\cite{Carloni}.
\\\\
\begin{figure}
\includegraphics[scale=0.8]{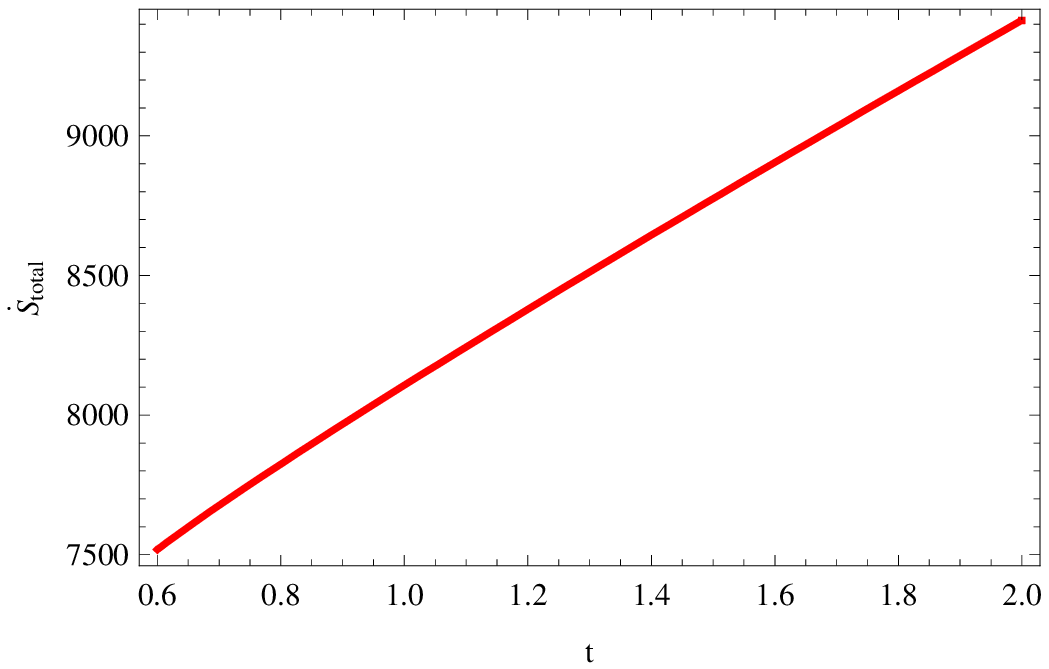}~~~~~\\
\vspace{2mm} \vspace{6mm} Fig. 1 plots $\dot{S}_{total}$ for
$\lambda=3, \mu=2$ under MFRHL considering the universe as a
thermodynamical system with the apparent horizon surface being its
boundary. \vspace{6mm}
\end{figure}

\begin{figure}
\includegraphics[scale=0.8]{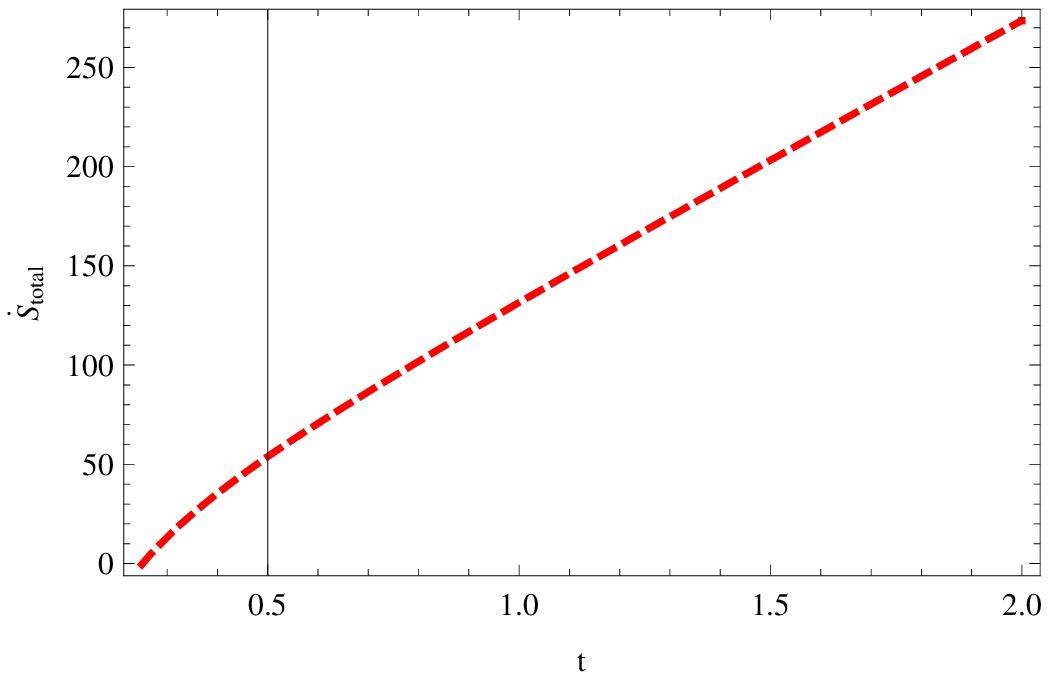}~~~~~\\
\vspace{2mm} \vspace{6mm} Fig. 2 plots $\dot{S}_{total}$ for
$\lambda=1, \mu=1$ under MFRHL considering the universe as a
thermodynamical system with the apparent horizon surface being its
boundary. This choice of $\lambda$ and $\mu$ corresponds to the
usual $f(R)$ gravity in spatially flat FRW cosmology.\vspace{6mm}
\end{figure}

\section{Modified $f(R)$ Horava-Lifshitz gravity}

An exhaustive review on Horava-Lifshitz (HL) cosmology is
available in the reference \cite{Mukhoyama}. There are a number of
cosmological implications of HL gravity. These are discussed in
the references \cite{Calcagni} and \cite{Brand}. The basic
quantities in HL gravity are: lapse $N(t)$, shift
$N^{i}(t,\overrightarrow{x})$ and 3D metric
$g_{ij}(t,\overrightarrow{x})$. All these terms are thoroughly
explained in the review entitled ``Horava-Lifshitz cosmology: a
review" published in \emph{Class. Quantum Grav.}, 27, 223101
(2010) by Mukhoyama \cite{Mukhoyama}. There is another review work
by Kiritsis and Kofinas \cite{Kiritsis} [\emph{Nuclear Physics},
821, 467 (2009)], where HL cosmology has been discussed. For the
dynamical variables mentioned above, the one can get a
four-dimensional metric as \cite{Mukhoyama},\cite{Kiritsis}
\begin{equation}
ds^{2}=-N^{2}dt^{2}+g_{ij}(dx^{i}+N^{i}dt)(dx^{j}+N^{j}dt)
\end{equation}
The lapse $N$ as stated above, is a function of time only.
However, the shift $N^{i}$ and $3$D metric $g_{ij}$ depend both on
time $t$ and spatial coordinate $\overrightarrow{x}$. The
condition on lapse is called the \emph{projectability condition}
\cite{Mukhoyama}.
\\
Recently an extension of the $f(R)$ gravity to a
Horava-Lifshitz-type theory has been proposed by \cite{Kluson}
introducing the action \cite{Chaichian}

\begin{equation}
S_{F_{HL}}=\int d^4x\sqrt{g^{(3)}}NF(R_{HL}),
~~~~~~~~~~~~~~R_{HL}\equiv K^{ij}K_{ij}-\lambda K^2-E^{ij}
{\cal{G}}_{ijkl}E^{kl}
\end{equation}
Here $\lambda$ is a real constant in the `generalised De Witt
metric' or `super-metric' (`metric of the space of metric')
\cite{Chaichian}
\begin{equation}
{\cal{G}}_{ijkl}=\frac{1}{2}(g^{(3)ik}g^{(3)jl}+g^{(3)il}g^{(3)jk})-\lambda
g^{(3)ij}g^{(3)kl}
\end{equation}
defined on three-dimensional hyperspace $\Sigma_t$, $E^{ij}$ can
be defined by the so-called detailed balanced condition by using
an action $W[g^3_{kl}]$ on the hypersurface $\Sigma_t$:
\begin{equation}
\sqrt{g^{(3)}}E^{ij}=\frac{\delta W[g^{(3)}_{kl}]}{\delta g_{ij}}
\end{equation}
and the inverse of ${\cal G}^{ijkl}$ is written as
\begin{equation}
{\cal{G}}_{ijkl}=\frac{1}{2}(g^{(3)}_{ik}g^{(3)}_{jl}+g^{(3)}_{il}g^{(3)}_{jk})-\tilde{\lambda}
g^{(3)}_{ij}g^{(3)}_{kl},
~~~~~~~~~~\tilde{\lambda}=\frac{\lambda}{3 \lambda-1}
\end{equation}

In the HL-like $f(R)$ gravity proposed by \cite{Kluson} as stated
above, the lapse $N$ was assumed to be a function of $t$ only,
which is the \emph{projectability condition}. Chaichian et al
\cite{Chaichian} proposed a new very general HL-like $f(R)$
gravity, which is a general approach for the construction of
modified gravity which is invariant under foliation-preserving
diffeomorphism. For the new form generalized gravity dubbed as
``modified $f(R)$ Horava-Lifshitz gravity" (MFRHL), reference
\cite{Chaichian} proposed the following action
\begin{equation}
S_{f(\tilde{R})}=\int d^{4}\sqrt{g^{(3)}}N f(\tilde{R})
\end{equation}
where,
\begin{equation}
\tilde{R}=K^{ij}K_{ij}-\lambda K^{2}+2\mu
\nabla_{\mu}\left(n^{\mu}\nabla_{\nu}n^{\nu}-n^{\nu}\nabla_{\nu}n^{\mu})-E^{ij}\mathcal{G}_{ijkl}E^{kl}\right)
\end{equation}
In flat FRW universe, the form of $\tilde{R}$ comes out to be

\begin{equation}
\tilde{R}=\frac{(3-9\lambda)H^2}{N^2}+\frac{6\mu}{a^3
N}\frac{d}{dt}\left(\frac{Ha^3}{N}\right)=\frac{\left(3-9\lambda+18\mu\right)H^2}{N^2}+\frac{6\mu}{N}\frac{d}{dt}\left(\frac{H}{N}\right)
\end{equation}

The $\tilde{R}$ of the above equation reduces to $R$ and
consequently the gravity reduces to usual $f(R)$ gravity for
$\lambda=\mu=1$ in a spatially flat FRW universe. For the action
in equation (10), we get by variation over $g_{ij}^{(3)}$ and by
setting $N=1$:

\begin{equation}
0=f(\tilde{R})-2(1-3\lambda+3\mu)(\dot{H}+3H^2)f'(\tilde{R})-2(1-3\lambda)H\frac{df'(\tilde
R)}{dt}+2\mu\frac{d^2f'(\tilde R)}{dt^2}+p
\end{equation}

In the above equation, $f'$ denotes the derivative with respect to
its argument. In the above equation, the matter contribution is
included by means of pressure $p$. Considering $\rho$ as the
matter density the conservation equation can be written as

\begin{equation}
\dot{\rho}+3H(\rho+p)=0
\end{equation}

and subsequently, equation (13) produces

\begin{equation}
0=f(\tilde{R})-6\left[(1-3\lambda+3\mu)H^{2}+\mu
\dot{H}\right]f'(\tilde{R})+6\mu
H\frac{df'(\tilde{R})}{dt}-\rho-Ca^{-3}
\end{equation}

where, $Ca^{-3}$ is regarded as dark matter. In the present work
we have considered $C\neq 0$. In this situation we have considered
both $\lambda=\mu=1$ as well as $\lambda\neq1,~~\mu\neq1$. It
should be further noted that we have considered the scale factor
$a$ in the power law form i.e. $a\propto t^{n}$ ($n$ is a positive
real number) as used earlier by the references like \cite{setare},
\cite{chatto1}, \cite{zimdahl} and \cite{sami}.\\
Purpose of the present work is to investigate the validity of the
generalized second law of thermodynamics (GSL) in the modified
gravity theory proposed by \cite{Chaichian}, i.e. in the MFRHL.
For the field equation of the form (13) we shall investigate
whether the total entropy of the universe i.e. sum of the time
derivatives of the entropy on the horizon and inside the horizon
stays at positive level. As the horizons for the enveloping
surfaces we have considered the cases of event horizon, apparent
horizon and particle horizons. Earlier \cite{pavon} explored the
thermodynamics of dark energy taking into account the existence of
the observer's event horizon in accelerated universes and showed
that except for the initial stage of Chaplygin gas dominated
expansion, the generalized second law of gravitational
thermodynamics is fulfilled. Jamil et al \cite{jamil} investigated
the validity of the GSL of thermodynamics in a universe governed
by Horava-Lifshitz gravity considering the universe as a
thermodynamical system bounded by the apparent horizon. Bamba and
Geng \cite{Bamba} investigated the GSL in $f(R)$ gravity with
realizing a crossing of the phantom divide in a universe enveloped
by the apparent horizon. Debnath et al \cite{chatto2} investigated
the GSL in various scenarios of the universe in the framework of
fractional action cosmology for apparent, event and particle
horizons. The issues associated with the GSL of thermodynamics are
thoroughly discussed in the references mentioned above. However,
for convenience we are giving a brief overview of the GSL of
thermodynamics in the subsequent section.
\\\\

\section{Generalized second law}
The connection between gravitation and thermodynamics was examined
by following black hole thermodynamics (black hole entropy
\cite{Bekenstein} and temperature \cite{hawking}) and its
application to the cosmological event horizon of de Sitter space
\cite{Gibbons}. Significant works are available on the study of
the generalized second law of thermodynamics in cosmology. The
studies include \cite{setare1}, \cite{setare2}, \cite{sheykhi},
\cite{jamil1}, \cite{jamil2} and \cite{debnath}. The basic
necessity for the validity of GSL is that the time derivative of
the total entropy $\dot{S}_{Total}=\dot{S}_{H}+\dot{S}\geq 0$,
where $\dot{S}$ indicates the time derivative of normal entropy
and $\dot{S}_{H}$ indicates the horizon entropy \cite{pavon}.
Thermodynamics under generalized gravity theories has been
discussed in \cite{Wu}.
\\
The first law of thermodynamics (Clausius relation) on the horizon
is defined as $T_{X}dS_{X}=\delta Q=-dE_{X}$. From the unified
first law, we may obtain the first law of thermodynamics as
\cite{samarpita}

\begin{equation}
T_{X}dS_{X}=4\pi R_{X}^{3}H (\rho_{eff}+p_{eff})dt
\end{equation}

where, $T_{X}$ and $R_{X}$ are the temperature and radius of the
horizons under consideration in the equilibrium thermodynamics.
The suffix $X$ will be replaced by $E$, $A$ and $P$ for event,
apparent and particle horizons respectively. Subsequently, the
time derivative of the entropy on the horizon $(\dot{S}_{X})$ and
inside the horizon $(\dot{S}_{IX})$ can be derived as
\cite{samarpita}

\begin{equation}
\dot{S}_{X}=\frac{4\pi R_{X}^{3}H}{T_{X}}(\rho_{eff}+p_{eff})~;~~
\dot{S}_{IX}=\frac{4\pi
R_{X}^{2}}{T_{X}}(\rho_{eff}+p_{eff})(\dot{R}_{X}-HR_{X})
\end{equation}

\begin{figure}
\includegraphics[scale=0.8]{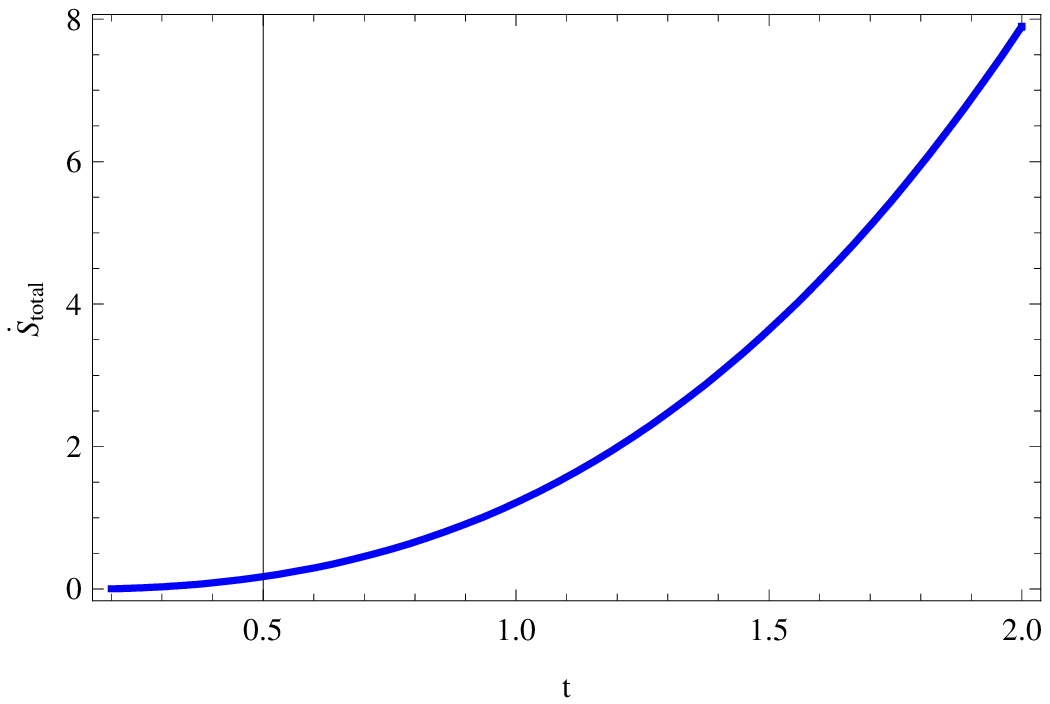}~~~~~\\
\vspace{2mm} \vspace{6mm} Fig. 3 plots $\dot{S}_{total}$ for
$\lambda=3, \mu=2$ under MFRHL considering the universe as a
thermodynamical system with the event horizon surface being its
boundary. \vspace{6mm}
\end{figure}

\begin{figure}
\includegraphics[scale=0.8]{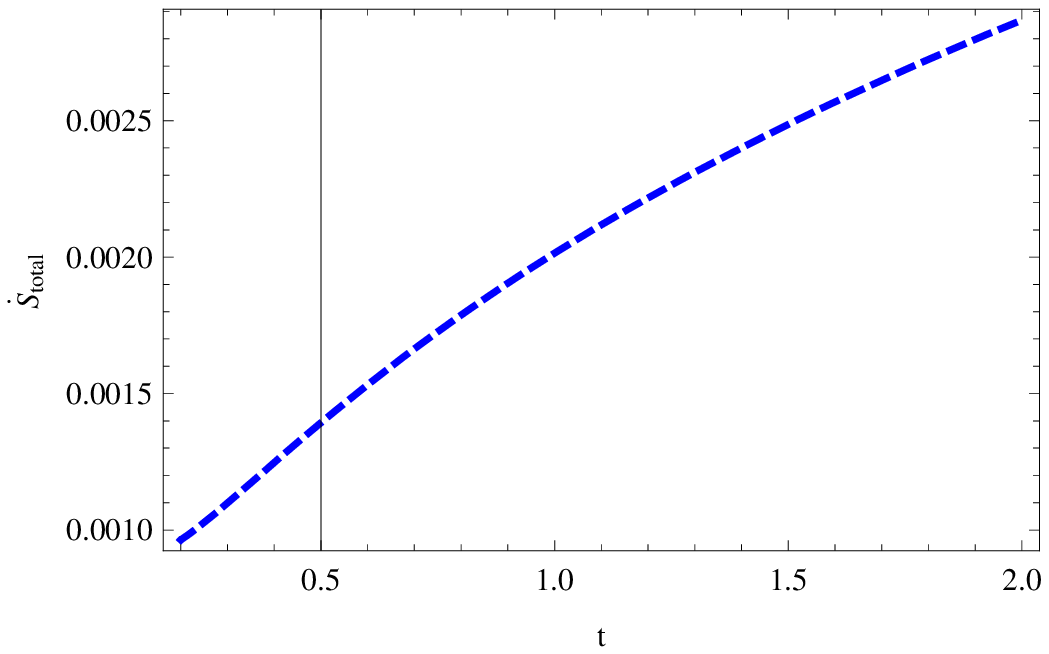}~~~~~\\
\vspace{2mm} \vspace{6mm} Fig. 4 plots $\dot{S}_{total}$ for
$\lambda=1, \mu=1$ under MFRHL considering the universe as a
thermodynamical system with the event horizon surface being its
boundary. This choice of $\lambda$ and $\mu$ corresponds to the
usual $f(R)$ gravity in spatially flat FRW cosmology. \vspace{6mm}
\end{figure}

Finally, we can get the time derivative of total entropy as
\cite{arundhati}

\begin{equation}
\dot{S}_{Total}=\dot{S}_{X}+\dot{S}_{IX}=\frac{
R_{X}^{2}}{GT_{X}}\left(\frac{k}{a^{2}}-\dot{H}\right)\dot{R}_{X}
\end{equation}

Our target is to investigate whether $\dot{S}_{X}+\dot{S}_{IX}\geq
0$ holds. The radii of apparent, event and particle and event
horizon are given by \cite{chatto2}

\begin{equation}
R_{A}=\frac{1}{\sqrt{H^{2}+\frac{k}{a^{2}}}}~;~~~
R_{E}=a\int_{t}^{\infty}\frac{dt}{a}~;~~~R_{P}=a\int_{0}^{t}\frac{dt}{a}
\end{equation}

In equation (18), we replace $X$ by $A$, $E$ and $P$ respectively
for investigating the GSL in MFRHL. It should be noted that as we
are considering MFRHL in flat FRW universe, we shall take $k=0$,
for which apparent horizon reduces to Hubble horizon. To study the
GSL of thermodynamics in MFRHL, we first consider $f(\tilde{R})$
as \cite{Nojiri3}
\begin{equation}
f(\tilde{R})=\tilde{R}+\xi\tilde{R}^{\eta}+\zeta \tilde{R}^{\nu}
\end{equation}

Here, $\xi,~\eta,~\zeta~\text{and}~\nu$ are real constants. Using
equations (13) and (15) we get the expressions for energy density
and pressure. From (14) we get the solution for dark matter as
$\rho=\rho_{m0}a^{-3}$ (i.e. $C=\rho_{m0}$ in (15)). Using the
expression for $\tilde{R}$ from equation (12) in equation (20) and
subsequently using (13), (15) and (18) we get the time derivatives
for total entropy for different horizons. For apparent horizon,
the time derivative of total entropy becomes
\begin{equation}
\begin{array}{c}
\dot{S}_{total}=-\frac{2\pi t^{1-3n}}{3Gn^{4}(n-3n\lambda-2\mu+6n\mu)}\times\\
\left[3n^{3}t^{3n}(-1+8\pi G(-1+3\lambda))(-1+3\lambda-6\mu)+16G\pi t^{2+3n}\mu(3^{\nu}\zeta(nt^{-2}(n-3n\lambda-2\mu+6n\mu))^{\nu})\right.\\
\left.\nu(1-3\nu+2\nu^{2})+3^{\eta}\eta(1-3\eta+2\eta^{2})(nt^{-2}(n-3n\lambda-2\mu+6n\mu))^{\eta}\xi\}\right]+\\
2n[3t^{3n}\mu-4G\pi t^{2+3n}(3^{\nu}\zeta(nt^{-2}(n-3n\lambda-2\mu+6n\mu))^{\nu}\nu(1-6\mu(-1+\nu)-\\
2\nu+\lambda(-3+6\nu))+3^{\eta}\eta(1-3\lambda+\eta(-2+6\lambda-6\mu)+6\mu)(nt^{-2}(n-3n\lambda-2\mu+6n\mu)^{\eta}\xi)-12G\pi t^{2}\mu\rho_{m0})+\\
 3n^{2}(t^{3n}(-1-8(1+2\pi G)\mu+\lambda(3+48\pi G\mu))+4\pi G t^{2}(1-3\lambda+6\mu)\rho_{m0})]\\
\end{array}
\end{equation}
In figure 1, we plot $\dot{S}_{total}$ against cosmic time $t$ for
$\lambda=3$ and $\mu=2$. For $\lambda=\mu=1$ we plot
$\dot{S}_{total}$ in figure 2. This case, as earlier mentioned,
corresponds to the usual $f(R)$ gravity. In both of the cases we
find the $\dot{S}_{total}$ to be in positive level. This confirms
the validity of GSL of thermodynamics for usual $f(R)$ as well as
modified $f(R)$ Horava-Lifshitz gravity proposed by
\cite{Chaichian}, where $R$ is replaced by $\tilde{R}$. To get the
$\dot{S}_{total}$ for event and particle horizons we adopt a
little different procedure. For these two horizons we consider the
derivatives
\begin{equation}
\dot{R}_{E}=HR_{E}-1~~;~~~\dot{R}_{P}=HR_{P}+1
\end{equation}

\begin{figure}
\includegraphics[scale=0.8]{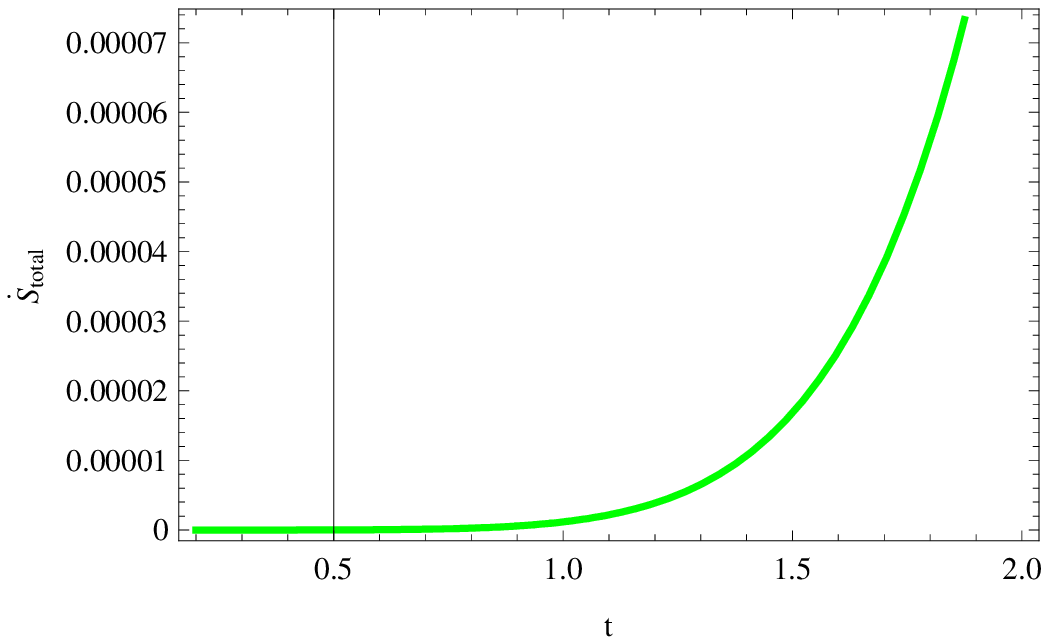}~~~~~\\
\vspace{2mm} \vspace{6mm} Fig. 5 plots $\dot{S}_{total}$ for
$\lambda=3, \mu=2$ under MFRHL considering the universe as a
thermodynamical system with the particle horizon surface being its
boundary. \vspace{6mm}
\end{figure}

\begin{figure}
\includegraphics[scale=0.8]{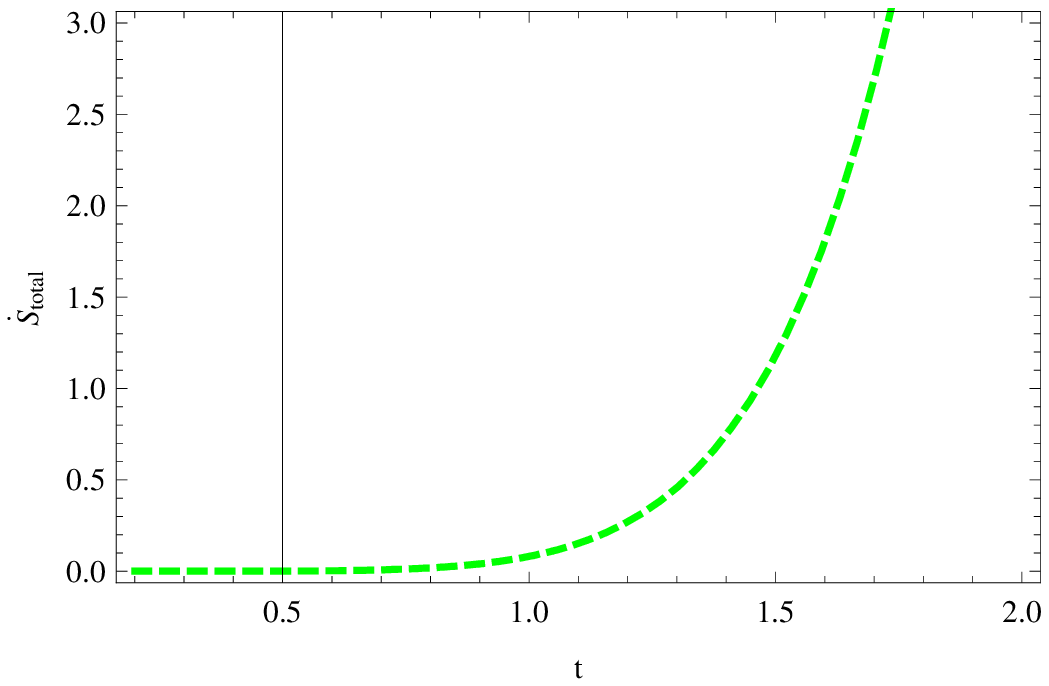}~~~~~\\
\vspace{2mm} \vspace{6mm} Fig. 6 plots $\dot{S}_{total}$ for
$\lambda=1, \mu=1$ under MFRHL considering the universe as a
thermodynamical system with the particle horizon surface being its
boundary. This choice of $\lambda$ and $\mu$ corresponds to the
usual $f(R)$ gravity in spatially flat FRW cosmology. \vspace{6mm}
\end{figure}

Using (22) in (18) we get the plots for $\dot{S}_{total}$
corresponding to event and particle horizons. For both of the
horizons, the $\dot{S}_{total}$ is found to stay at positive level
irrespective of the values of $\lambda$ and $\mu$. Finally, before
concluding we consider the equation of state parameter for the
MFRHL.

\begin{figure}
\includegraphics[scale=0.8]{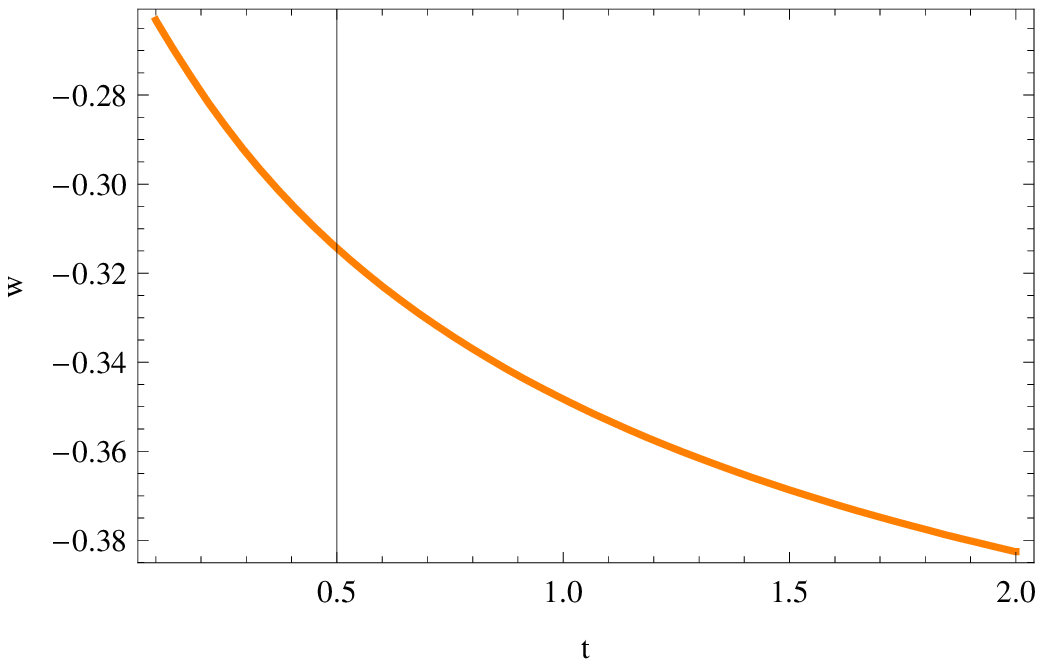}~~~~~\\
\vspace{2mm} \vspace{6mm} Fig. 7 plots equation of state parameter
$w$ for $\lambda=3, \mu=2$ under MFRHL. Here $w>-1$ that indicates
quintessence like behaviour. \vspace{6mm}
\end{figure}

\begin{figure}
\includegraphics[scale=0.8]{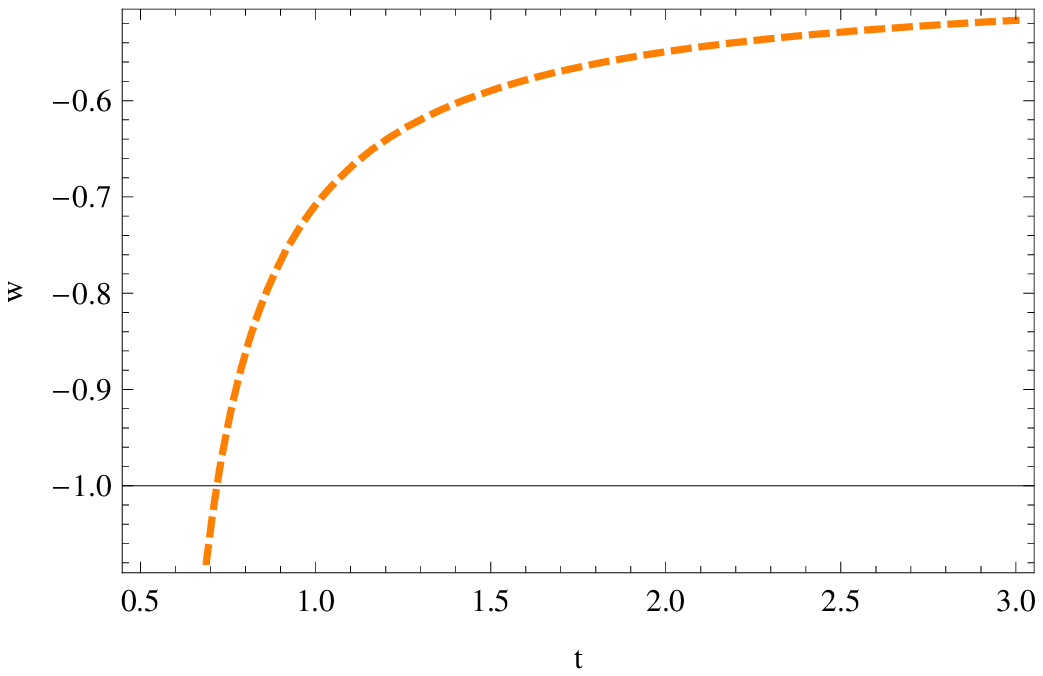}~~~~~\\
\vspace{2mm} \vspace{6mm} Fig. 8 plots equation of state parameter
$w$ for $\lambda=1, \mu=1$ under MFRHL. It crosses the phantom
divide $w=-1$ and hence behaves like quintom \cite{Saridakis2}.
This choices of $\lambda$ and $\mu$ correspond to the usual $f(R)$
gravity in spatially flat FRW cosmology. \vspace{6mm}
\end{figure}

For $\lambda=3,\mu=2$, we plot the equation of state parameter in
figure 7, where it is observed that the equation of state
parameter lies above $-1$. This indicates that the equation of
state parameter behaves like quintessence. However, in figure 8,
(where $\lambda=\mu=1$) we plot the equation of state parameter is
crossing the phantom divide $-1$. It has been already stated that
for $\lambda=\mu=1$ the modified $f(R)$ Horava-Lifshitz gravity
reduces to usual $f(R)$ gravity and reference \cite{Bamba} proved
that under usual $f(R)$ gravity the phantom divide is crossed.
\\\\
\section{Concluding remarks}
We have investigated the generalized second law of thermodynamics
in modified $f(R)$ Horava-Lifshitz gravity proposed by
\cite{Chaichian}. In reference \cite{Carloni} modified
Horava-Lifshitz $f(R)$ gravity was considered for barotropic fluid
and it was shown to have a quite rich cosmological structure:
early/late-time cosmic acceleration of quintessence, as well as of
phantom types. It has been shown in reference \cite{Chaichian}
that for $\lambda=\mu=1$, the modified $f(R)$ Horava-Lifshitz
gravity reduces to usual $f(R)$ gravity. We have examined the
equation of state parameter $w$ for $\lambda=3,~\mu=2$ i.e. for
modified $f(R)$ Horava-Lifshitz gravity. It is observed that $w$
is staying above $-1$ that indicates quintessence
\cite{Saridakis2}. Thus, one notable difference between $f(R)$ and
modified $f(R)$ Horava-Lifshitz gravity is that the second one
does not realize the phantom divide, when the scale factor is
taken in power law form $(a\propto t^{n})$. Rather its equation of
state behaves like quintessence. The generalized second law of
thermodynamics is satisfied not only in $f(R)$ gravity, but also
in modified $f(R)$ Horava-Lifshitz gravity. Moreover, the radii of
the enveloping horizon does not affect the validity of the
generalized second law. Irrespective of the radius of the
enveloping horizon, the time derivative of total entropy is
staying at positive level and is increasing with the evolution of
the universe. \\
While concluding the paper we would like to state that although we
have considered the power-law form of scale factor in this work,
other forms like logamediate, intermediate, emergent etc. Scale
factor are possible also via using reconstruction scheme
\cite{Odintsov}. We propose to investigate the thermodynamic laws
in the modified $f(R)$ Horava-Lifshitz gravity for such choices
too in future works.
\\\\
\subsection{Acknowledgment}
The first author dedicates this work to the loving memory of his
mother Kamala Chattopadhyay, who was the source of all
inspirations in his life. The first author sincerely acknowledges
the Inter-University Center for Astronomy and Astrophysics
(IUCAA), Pune, India, which provided Visiting Associateship to the
author for the period of August 2011 to July 2014. The second
author wishes to thank IUCAA, Pune for providing all facilities to
carry out this research during a scientific visit in January,
2012. Sincere thanks are due to the anonymous reviewer for
constructive comments on this work.
\\


\begin{thebibliography}{100}
\bibitem{Riess} A.G. Riess et al., Astron. J. \textbf{116} (1998) 1009.
\bibitem{padmanabhan}T. Padmanabhan, Curr. Sci. \textbf{88} (2005)  1057.
\bibitem{Nojiri} S. Nojiri S and S. D. Odintsov, Int. J. Geom. Methods Mod. Phys. \textbf{4} (2007) 115.
\bibitem{Saridakis} E. N. Saridakis, Eur. Phys. J. C \textbf{67} (2010) 229.
\bibitem{Chaichian} M. Chaichian, S. Nojiri, S. D. Odintsov, M.
Oksanen, and A. Tureanu, Class. Quantum Grav. \textbf{27} (2010)
185021.
\bibitem{Gourgoulhon} E. Gourgoulhon, arXiv:gr-qc/0703035
(2007).
\bibitem{Kluson} J. Kluson, Phys. Rev. D \textbf{81}(2009) 064028.
\bibitem{Bekenstein} J. D. Bekenstein, Phys. Rev. D \textbf{7} (1973) 2333.
\bibitem{hawking}  S. W. Hawking, Commun. Math. Phys. \textbf{43} (1975)
199.
\bibitem{Gibbons} G. W. Gibbons, S. W. Hawking, Phys. Rev. D \textbf{15} (1977) 2738.
\bibitem{Bamba} K. Bamba, C-Q. Geng, Physics Letters B \textbf{679} (2009)
282.
\bibitem{Carloni} S. Carloni et al., Phys. Rev. D \textbf{82}
(2010) 065020.
\bibitem{Woodard} R. P. Woodard, Lect. Notes Phys. \textbf{720} (2007)
403.
\bibitem{Mukhoyama} S. Mukhoyama, Class. Quant. Grav., \textbf{27}
(2010) 223101.
\bibitem{Calcagni} G. Calcagni, J. High Energy Phys. \textbf{09}
(2009) 112.
\bibitem{Brand} R. Brandenberger, Phys. Rev. D \textbf{80} (2009)
043516.
\bibitem{Kiritsis} E. Kiritsis and G. Kofinas, Nuclear Physics
\textbf{821} (2009) 467.
\bibitem{setare} A. R. Rastkar, M. R. Setare, F. Darabi,
Astrophys. Space Sci. (online first 2011), DOI
10.1007/s10509-011-0849-9
\bibitem{chatto1} S. Chattopadhyay, U. Debnath, Int. J Theor. Phys. \textbf{49} (2010)
1465.
\bibitem{zimdahl} W. Zimdahl, D. Pavón, L. P. Chimento , Phys. Lett. B \textbf{521} (2001)
133.
\bibitem{sami} E. J. Copeland, M. Sami, S. Tsujikawa, Int. J. Mod. Phys. D \textbf{15} (2006)
1753.
\bibitem{pavon} G. Izquierdo and D. Pavón, Phys. Lett. B \textbf{633} (2006)
420.
\bibitem{Wu} S-F. Wu, B. Wang and G-H. Yang, Nuclear Physics B \textbf{799} (2008)
330.
\bibitem{samarpita} S. Bhattacharya and U. Debnath, Canadian
Journal of Physics \textbf{89} (2011) 883.
\bibitem{jamil} M. Jamil, E. N. Saridakis, M. R. Setare, JCAP \textbf{11} (2010)
032.
\bibitem{chatto2} U. Debnath, M. Jamil, S. Chattopadhyay, Int. J Theor.
Phys.(Online first, 2011) DOI 10.1007/s10773-011-0961-1.
\bibitem{arundhati} A. Das, S. Chattopadhyay, U. Debnath, Foundations of Physics (Online first, 2011) DOI
10.1007/s10701-011-9600-1.
\bibitem{Nojiri3} S. Nojiri and S. D. Odintsov , Phys. Rev.
D \textbf{68} (2003) 123512.
\bibitem{Saridakis2} Y-F. Cai, E. N. Saridakis, M. R. Setare and
J-Q. Xia, Physics Reports \textbf{493} (2010) 1.
\bibitem{setare1} M. R. Setare, Phys. Lett. B \textbf{641} (2006)
130.
\bibitem{setare2} M. R. Setare, JCAP \textbf{0701} (2007) 023.
\bibitem{jamil1} M. Jamil, I. Hussain, Int. J. Theor. Phys. \textbf{50} (2011)
465.
\bibitem{jamil2} M. Jamil and M. Akbar, General Relativity and
Gravitation, \textbf{43} (2011) 1061.
\bibitem{debnath} U. Debnath, EPL \textbf{94} (2011) 29001.
\bibitem{sheykhi} A. Sheykhi, Class. Quantum Grav.\textbf{27} (2010)
025007.
\bibitem{Carloni} S. Carloni, M. Chaichian, S.
Nojiri, S. D. Odintsov, M. Oksanen and A. Tureanu, Phys. Rev. D
\textbf{82} (2010) 065020.
\bibitem{Odintsov} S. Nojiri, S. D. Odintsov, Phys. Rept. \textbf{505} (2011)
59.

\end{thebibliography}
\end{document}